\documentclass[a4paper,11pt, draftcls, onecolumn]{IEEEtran}
\usepackage[latin1]{inputenc}
\usepackage[english]{babel}
\usepackage[T1]{fontenc}
\usepackage{array}
\usepackage{graphicx}
\usepackage[caption=false,font=footnotesize]{subfig}
\begingroup\expandafter\expandafter\expandafter\endgroup
\expandafter\ifx\csname IncludeInRelease\endcsname\relax
  \usepackage{fixltx2e}
\fi
\usepackage{cite}
\usepackage{algorithm}              
\usepackage{algorithmic}
\usepackage{xfrac}
\usepackage{wrapfig}    
\usepackage[cmex10]{amsmath}
\usepackage{amsfonts,amssymb}
\usepackage{booktabs}
\usepackage{color}
\usepackage{epstopdf}
\usepackage{float}
\usepackage[T1]{fontenc}
\usepackage[cmex10]{amsmath}
\usepackage{amsfonts,amssymb}
\usepackage{multicol}
\usepackage{url}
\usepackage{yfonts}

\usepackage{tikz}
\definecolor{gold}{rgb}{0.85,.66,0}
\definecolor{cian}{rgb}{.02,.7,.95}
\definecolor{ppp}{rgb}{.7,.3,.82}
\definecolor{red}{rgb}{1,0,0}
\begin{document}

\title{Sequential Likelihood Ascent Search Detector for Massive MIMO Systems}
\author{Giovanni~Maciel~Ferreira~Silva \& Jos\'e~Carlos~Marinello~Filho \& Taufik~Abr\~ao
\thanks{State University of Londrina, Electrical Engineering Department (DEEL-UEL).  \\
 Rod. Celso Garcia Cid - PR445,   s/n,    Campus Universitario, 
         Po.Box 10.011; \quad  86057-970, \\ Londrina, \quad 
        Parana  - Brazil; \quad Phone:   +55 43 3371 4790\\
E-mails: {giomaciel.fs@gmail.com}, {zecarlos.ee@gmail.com}, {taufik@uel.br}}
}

\maketitle

\begin{abstract}
In this paper, we have analyzed the performance-complexity tradeoff of {a selective} likelihood ascent search (LAS) algorithm initialized by a linear detector, such as matched filtering (MF), zero forcing (ZF) and minimum mean square error (MMSE), {and considering an optimization factor $\rho$ from the bit flipping rule}. The scenario is the uplink of a massive MIMO (M-MIMO) system, and the {analysis has }been developed by means of computer simulations. With the increasing number of base station (BS) antennas, the classical detectors become inefficient. Therefore, the LAS is employed for performance-complexity tradeoff improvement. Using an adjustable optimized threshold on the bit flip rule of LAS, much better solutions have been achieved in terms of BER with no further complexity increment, {indicating that there is an optimal threshold for each scenario. Considering a $32\times32$ antennas scenario, {the large-scale MIMO system eqquiped with the proposed LAS  detector with} factor $\rho$ = 0.8 {requires} 5 dB less {in terms of SNR than the conventional LAS of the} literature ($\rho$ = 1.0) to achieve the same bit error rate of $10^{-3}$.}
\end{abstract}

\begin{IEEEkeywords}
Massive MIMO; likelihood ascent search;  linear detector; threshold analysis.
\end{IEEEkeywords}

%-------------------------------------------------------------------------------
\section{Introduction}\label{sec:intro}
%-------------------------------------------------------------------------------
The next generation of wireless communication systems (5G) aims to deliver low latency, high data rates combined to high reliability \cite{Agiwal_5G}. A promising research area in 5G is massive MIMO (M-MIMO) systems. M-MIMO is an emerging telecommunications technology expected to integrate the fifth generation (5G) systems standards around 2020 \cite{Rupam_5G}. In contrast to conventional MIMO systems, a Massive MIMO system can feature hundreds of antennas. Increasing the number of antennas brings some advantages such as the channel hardening effect \cite{LargeMIMOSystems}, in addition to reliability, energy and spectral efficiencies. Therefore, efficient processing techniques on the transmitter and receiver/detector are required.

An M-MIMO system allows linear operations in large scale, such as precoding and decoding \cite{Marzetta_MIMO2015}. In general, systems work in spatial modulation or spatial diversity. To increase data rates, spatial modulation is often used, which depends on a good knowledge of the communication channel \cite{Marzetta_NextGeneration2014}. Assuming that the receiver has perfect knowledge of the channel state information (CSI), it is possible to accurately detect the information. For this, it is necessary that there is processing for both channel estimation and information detection. As the number of antennas increases, there is an increase in complexity in both processes. Conventional MIMO detectors perform almost optimally when the number of transmitting antennas is much larger than the number of receiving antennas ($N_t >> N_r$). However, with increasing numbers of users, scenarios where $N_t \approx N_r$ are also practical in M-MIMO \cite{Rusek_13}.  In \cite{Marzetta_MIMO2015}, the author analyzed a scenario with 64 BS antennas serving 18 single-antenna users. So, $N_t = N_r$ can be considered a limit condition  to verify the ultimate detection performance in a macrocell uplink M-MIMO system.

The focus of this paper is on the data detection at the receiving antennas. Nonlinear detectors, such as maximum likelihood (ML) and sphere decoding (SD), are not feasible for use in M-MIMO systems, because their complexities increase significantly with the number of base station (BS) antennas. Linear detectors, such as matched filtering (MF) and zero forcing (ZF), present low complexity, but they also present a unacceptable performance with the increasing number of antennas. In M-MIMO scenarios, it is very important to have linear processing \cite{Marzeta_Myths2016} due to the large numbers of users and the low time available for processing. Thus linear detection techniques play an important role in the initial estimation of the transmitted information. Therefore, novel detection schemes aiming to achieve better performance-complexity tradeoffs are investigated in this paper.

The local search (LS) algorithms have been proposed in the context of multiuser detection for direct sequence/code division multiple access (DS/CDMA)  in \cite{Oliveira.2009}. More recently, low-complexity search subspace procedures for M-MIMO detection have been explored, {\it e.g.}, deploying heuristic approaches, such as layered tabu search algorithm for large MIMO detection \cite{Karthikeyan.2018}. They iteratively search for the vector which minimizes the ML cost in a fixed or reduncing neighborhood subspace. However, the ML solution may not lie in the reduced search subspace while the search may go through a large number of intermediate vectors.  The optimization procedure described in this work is associated to the likelihood ascent search (LAS) detector for M-MIMO system, which consists in iteratively improving the likelihood function of the ML detector with affordable complexity increment. Although detection through ascending likelihood search has been previously studied in CDMA systems \cite{Sun2000}, \cite{Sun1998}, recently there are a lot of interest in developing optimization procedures of the low-complexity quasi-optimum detectors directly applicable to the massive MIMO systems, such {as} those found in \cite{linbo2016}, \cite{sah2016}, \cite{Sah2017}. 

In the sequential and global LAS detection for M-MIMO proposed in \cite{Sah2017}, first it is cut down the size of the neighborhood, reducing the  algorithm complexity; secondly, the updating is not restricted to be in a fixed neighborhood, hence improving the detection performance. For the first purpose, authors in \cite{Sah2017} propose a metric and a few selection rules to decide whether or not to include a vector in the neighborhood. The indices of the largest components of the metric is deployed for generating a reduced neighborhood set, reducing the complexity of the existing algorithms while maintaining their error performance. {While \cite{Sah2017} changes the neighborhood set of SLAS and focuses on the complexity reduction, the central contribution in the proposed detector consists of improving performance with a marginal complexity increasing by changing the bit flip rule.}

The  LAS detection performance can be improved via optimization techniques, as in \cite{Chihaoui_2016}, where the antenna selection is made before the detector acts,  aiming at maximizing the SNR and capacity. In  \cite{Sah_2017}, the search for the best solution is done in a non-fixed neighborhood. In \cite{Li_2010}, an algorithm that uses multiple outputs is developed. {Differently from \cite{Chihaoui_2016}, \cite{Sah_2017}, \cite{Li_2010}, which consider some ways to improve the bit error rate (BER) performance based on the likelihood ascent search while increasing complexity, our approach reveals that there is an optimal value for the bit flip parameter. As a result, this approach improves the BER performance with a marginal complexity increase.}

{The {\it contribution} of this work is threefold. {\bf i}) an adjustable bit flipping rule threshold value SLAS M-MIMO detection scheme operating under an improved performance-complexity tradeoff is proposed; {\bf ii}) we have proposed and characterized the suitable threshold factor value choice for the bit flipping rule of LAS as a function of the number of antennas and SNR region. Hence, the appropriate choice of the threshold factor value can improve the BER performance substantially, differing from the proposed in \cite{Vardhan_LAS08}. Indeed, there is an optimal threshold factor for each system configuration. {\bf iii}) In M-MIMO systems, the number of antennas grows significantly. Hence, we have proposed an adaptation for the SLAS detector dealing with such requirement; as a consequence, the proposed SLAS M-MIMO detector can improve remarkably the BER performance, with only a marginal complexity increasing.}

The remainder of this paper is organized as follows. The M-MIMO system model used in this work is presented in Section \ref{sec:model}. Section \ref{sec:optimization} describes how the LAS optimization is performed after an initial linear detection. The numerical simulation analysis, including convergence, threshold analysis and complexity, are discussed in Sections \ref{sec:results} and \ref{sec:complexity}, respectively. Final remarks and conclusions are driven in Section \ref{sec:concl}.

%\vspace{-2.5mm}

%-------------------------------------------------------------------------------
\section{System Model}\label{sec:model}
%-------------------------------------------------------------------------------
%Consider a {M-MIMO uplink} system with $N_t$ {users equidistant} from $N_r$ receiving antennas {in the base station. As the distance is equal among users, it is possible to consider only the small-scale fading since the distance affects all users in the same way. The long-term fading is included in the signal-to-noise ratio (SNR) variation.}
{Consider a M-MIMO uplink system with $N_t$ user equipments (UEs) apart from $N_r$ receiving antennas in the base station (BS). For simplicity, let's assume that power allocation step has been previously applied in such a way that all UE signals are receiving with equal power at the BS receiver. Hence, it is possible to consider only the small-scale fading effect since the large-scale (path-loss) channel term has been compensated by allocating different UEs transmitting power, as inversely proportional to UE-BS distance. Besides, the long-term fading effect can be included in the signal-to-noise ratio (SNR) level.}

The symbol $b_j$ is transmitted by the $j$th antenna. This symbol is multiplied by the path gain $h_{kj}$, which is complex Gaussian (zero mean and unit variance) and refers to the $j$th transmitting antenna and the $k$th receiving antenna. The white Gaussian noise $n_k$ has variance $N_0$ and is added to information immediately before the receiving antennas. Thus, the symbol $y_k$ received on the $k$th antenna is given by
\begin{equation}
\label{eq:modelo_bit}
y_k = \displaystyle\sum_{j=1}^{N_t} h_{kj}b_j + n_k.
\end{equation}
Considering all the transmitting and receiving antennas, it is possible to rewrite Eq. \eqref{eq:modelo_bit} in the matrix notation, such that
\begin{equation}
\mathbf{y} = \mathbf{H}\mathbf{b} + \mathbf{n}
\end{equation}
where $\mathbf{y}$ is the $N_r$$\times$1 received signal, $\mathbf{b}$ is the $N_t$$\times$1 transmitted signal, $\mathbf{H}$ is the $N_r$$\times$$N_t$ channel gain matrix and $\mathbf{n}$ is the $N_r$$\times$1 noise vector.

In this work,  the massive MIMO channels may not always result harden, since we have adopted a number of antennas not so high in certain scenarios. The channel hardening phenomena can be defined as follows. We specifically define {\it channel hardening} for independent Rayleigh fading, a type of non-line-of-sight (NLOS) channel implicitly assumed in path gain distribution in eq. \eqref{eq:modelo_bit}. Channel hardening is a phenomenon where the norms of the channel vectors $\{{\bf h}_k\}, \,\, k = 1, \ldots, N_r$, fluctuate only little. Hence, the propagation achieves channel hardening condition if \cite{Ngo.2016}:
\begin{equation}
\frac{||{\bf h}_k||^2}{\mathbb{E}\left[||{\bf h}_k||^2\right]} \rightarrow 1, \quad \text{as} \quad N_t  \rightarrow \infty, \quad \,\, k = 1, \ldots, N_r
\end{equation}

Some of the simplest linear techniques for detecting the transmitted signal are MF, ZF and MMSE, as long as it is assumed a perfect knowledge of the CSI at the receiver. {The equations from the estimated vector after the MF, ZF and MMSE detectors are}
\begin{equation}\label{eq:MF}
\mathbf{\widehat{b}}_\textsc{mf} = \mathbf{H}^H\mathbf{y},
\end{equation}
\begin{equation}\label{eq:ZF}
\mathbf{\widehat{b}}_{\textsc{zf}} = (\mathbf{H}^H\mathbf{H})^{-1}\mathbf{H}^H\mathbf{y},
\end{equation}
and
\begin{equation}\label{eq:MMSE}
\mathbf{\widehat{b}}_\textsc{mmse} = \left(\mathbf{H}^H\mathbf{H}+  \dfrac{N_0}{E_s}\mathbf{I}\right)^{-1}\mathbf{H}^H\mathbf{y}.
\end{equation}
where $E_s$ is the symbol energy and $N_0$ is the noise variance $N_0$. More details of each detector can be found in \cite{LargeMIMOSystems}.

When the number of antennas increases to tens or hundreds, {\it i.e.}, massive MIMO scenarios, linear detectors lose performance. Therefore, some improvements are required to maintain suitable performance without adding too much complexity. In this context, the LAS algorithm arises.

%-------------------------------------------------------------------------------
\section{LAS Algorithm}\label{sec:optimization}
%-------------------------------------------------------------------------------
The optimization in LAS is based on the ML principle, where in the context of MIMO detection the likelihood function $\Lambda(\mathbf{b})$ is given by \cite{Verdu}:
\begin{equation}
\label{eq:lamb_b}
\Lambda(\mathbf{b}) = \mathbf{b}^T\mathbf{H}^H\mathbf{y} + \mathbf{b}^T(\mathbf{H}^H\mathbf{y})^* - \mathbf{b}^T\mathbf{H}^H\mathbf{H}\mathbf{b}.
\end{equation}
Basically, we look for a vector $\mathbf{b}(n)$, updating it in $n$ iterations such that the likelihood function grows with each step.  Hence, it is convenient to deal with the likelihood function different on consecutive iterations: 
\begin{equation}
\Delta\Lambda(\mathbf{b}) = \Lambda(\mathbf{b}(n+1)) - \Lambda(\mathbf{b}(n)) \geq 0.
\end{equation}
The output of a linear detector (MF, ZF or MMSE) is deployed at the beginning of algorithm, so it is identified as step zero detection output,  $\mathbf{b}(0)$. Hence, the best candidate vector $\mathbf{b}$ is the one maximizing likelihood function, Eq.  \eqref{eq:lamb_b}.%\cite{Verdu}:
%
%\begin{equation}
%\label{eq:lamb_b}
%\Lambda(\mathbf{b}) = \mathbf{b}^T\mathbf{H}^H\mathbf{y} + \mathbf{b}^T(\mathbf{H}^H\mathbf{y})^* - \mathbf{b}^T\mathbf{H}^H\mathbf{H}\mathbf{b}.
%\end{equation}

In \cite{Vardhan_LAS08}, it is shown that, defining
\begin{equation}
\label{eq:yeff}
\mathbf{y}_{\rm eff} = \mathbf{H}^H\mathbf{y} + (\mathbf{H}^H\mathbf{y})^*,
\end{equation}
\begin{equation}
\label{eq:heff}
\mathbf{H}_{\rm eff} = \mathbf{H}^H\mathbf{H},
\end{equation}
\noindent and
\begin{equation}
\label{eq:hreal}
\mathbf{H}_{\rm real} = 2\mathbb{R}(\mathbf{H}_{\rm eff}),
\end{equation}
Hence, we can find the gradient simply:
\begin{equation}
\label{eq:grad}
\mathbf{g} = \mathbf{y}_{\rm eff} - \mathbf{H}_{\rm real}\mathbf{b}.
\end{equation}
Moreover, the second derivative of the likelihood function with respect to $\mathbf{b}$ is
\begin{equation}
\dfrac{\partial^2 (\Lambda (\mathbf{b}))}{\partial (\mathbf{b})^2} = -\mathbf{H}_{\rm real},
\end{equation}
\noindent which indicates a maximum or a minimum point on the neighborhood local search. There is a huge family of LAS detectors, quoted in \cite{Sun2000},\cite{Sun1998}, depending on the index vector of bit flip candidates at step $n$, denoted by $\mathbf{l}(n) \subseteq \{1,2,..,N_t\}$. Considering BPSK modulation and $\mathbf{b}(0) \in \{-1,+1\}^{N_t}$, the updated vector $\mathbf{b}(n+1)$ can be written as
\begin{equation}
\label{eq:bit_update}
\mathbf{b}(n+1) = \mathbf{b}(n) -2\sum_{i \in \mathbf{l}(n)} b_i(n)\mathbf{e}_i
\end{equation}
and the updated gradient $\mathbf{g}(n+1)$, following Eq. \eqref{eq:grad} and \eqref{eq:bit_update}, can be written as
\begin{equation}
\mathbf{g}(n+1) = \mathbf{g}(n) + 2\sum_{i \in \mathbf{l}(n)}b_i(n)(\mathbf{H}_{\rm real})_i
\end{equation}

In \cite{Sun2000},\cite{Sun1998}, it is proved that if the gradient entry of the $j$th antenna, namely $g_j$, exceeds the threshold $\zeta_j$, it is guaranteed that the cost function always increases. This threshold is given by the absolute value of the second derivative of likelihood function, given by
\begin{equation}
\label{eq:threshold_glas}
\zeta_j = \sum_{i \in \mathbf{l}(n)}\rho\;|\left(\mathbf{H}_{\rm real}\right)_{ji}|,\;\forall j \in \mathbf{l}(n).
\end{equation}
{where $\rho$ is the selective adjustable factor that can smooth or harden the bit flipping rule, depending on the number of antennas and SNR level. In \cite{Vardhan_LAS08}, $\rho$ is always unitary, but it can raise the performance in some M-MIMO scenarios.}

We have considered the sequential LAS (SLAS) proposed in \cite{Sah2017}. In SLAS, only one bit is flipped at each step, it is done circularly and vector $\mathbf{l}(n)$ has only one index per step. So, the threshold $\zeta_j$ for SLAS, specially in the context of massive MIMO detection, is constant and given rewriting Eq. \eqref{eq:threshold_glas} as
\begin{equation}\label{eq:threshold}
\zeta_j = \rho\;|(\mathbf{H}_{\rm real})_{j,j}|.
\end{equation}
%\colr{justificar por que no  sequential LAS o limiar Ã© $\zeta_j$  cte e dado pelos elementos da diagonal principal $(\mathbf{H}_{\rm real})_{j,j}$ !!!}

Our SLAS-based M-MIMO detector runs through all the bits of a symbol and updates them, always ensuring that there is a growth of the likelihood function value at the current iteration. For the case of BPSK, the update rule for the bit $b_j$ is given by %\cite{Vardhan_LAS08}:
\begin{equation}\label{eq:rule}
b_j(n+1) =
\begin{cases}
+1, \; \text{if } b_j(n) = -1 \text{ and } g_j(n) > \rho \; \zeta_j(n), \\
-1, \; \text{if } b_j(n) = +1 \text{ and } g_j(n) < -\rho \; \zeta_j(n), \\
b_j(n), \; \text{otherwise}.\\
\end{cases}
\end{equation}

The updating is done until a fixed number of iterations $n_{\textsc{f}}$ is reached. When $n$ = $n_{\textsc{f}}$, the algorithm ends and $\mathbf{b}(n_{\textsc{f}})$ is the final data vector estimated by the LAS algorithm. Since the algorithm is circular, if the number of steps $n$ is greater than the number of symbols contained in the information vector, {\it i.e.}, $n > \ell \cdot N_t$, where $\ell$ is the constellation number of bits, then the algorithm returns to the first position and continues to run, performing more than one sequential iteration per antenna. A pseudo-code for the Selective-$\rho$ SLAS M-MIMO detector is depicted in Algorithm \ref{alg:LAS}.
\linespread{1.1}
\begin{algorithm}[!htbp]
	\caption{{Selective-$\rho$} Sequential LAS M-MIMO detector}\label{alg:LAS}
	\begin{algorithmic}[1]
		\STATE \textit{{Inicialize}}
		\STATE $\mathbf{b}(0) \leftarrow$ \textit{MF, ZF or MMSE} decision vector
		\STATE $j$ = 1, $n$ = 0
		\STATE \textit{Define} $n_F$
		\WHILE{$n \leq n_F$}
		\IF{$j > N_t$}
		\STATE $j \leftarrow 1$ 
		\ELSE
		\STATE \textit{Calculate} $\mathbf{g}(n)$
		\STATE \textit{Calculate threshold} $\zeta_j$
		\IF{$b_j(n) = -1$ and $g_j(n) > {\rho} \; \zeta_j$}
		\STATE $b_j(n+1) \leftarrow +1$
		\ELSIF{$b_j(n) = +1$ and $g_j(n) < -{\rho} \; \zeta_j$}
		\STATE $b_j(n+1) \leftarrow -1$
		\ELSE
		\STATE $b_j(n+1) \leftarrow b_j(n)$
		\ENDIF
		\STATE $j \leftarrow j + 1$
		\STATE $n \leftarrow n + 1$
		\ENDIF
		\ENDWHILE
		\STATE \textit{Solution:} $\mathbf{b}(n_F)$
		\STATE \textit{{End}} 
	\end{algorithmic}
\end{algorithm}

%-------------------------------------------------------------------------------
\section{Numerical Results -- Performance and Threshold Optimization}\label{sec:results}
%-------------------------------------------------------------------------------
Numerical Monte-Carlo simulation (MCS) results have been analysed and classified in three groups: 
\begin{itemize}
	\item[{\bf a})] the performance of the proposed SLAS-based MIMO algorithm changing the SNR and the number of antennas; 
	\item[{\bf b})]  the likelihood and BER behavior at each step of the SLAS algorithm; 
	\item[{\bf c})]  the threshold $\zeta_j$ value choice related to the bit flipping rule in eq. \eqref{eq:rule}, which intrinsically determine the LAS performance.
\end{itemize}

MCS was extensively used considering  number of realizations in the range of  $\{10^3;\,\, 10^5\}$ trials aiming at assuring at least 5 bit errors on each simulated condition. Random information, additive noise, and short-term fading samples have been generated for each single MCS realization. On the other hand, the long- and medium-term, i.e., path-loss and shadowing, respectively, were included in the SNR level. In terms of system configuration, BPSK modulation, sequential LAS and a wide range of number of antennas at the receiver and transmitter sides were considered aiming at exploring M-MIMO scenarios. Moreover, the output vector for the first iteration $\mathbf{b}(0)$ has been obtained from the output of a linear detector, especially from the MF, ZF or alternatively the MMSE MIMO detectors. Tab. \ref{tab:parameters} indicates the main parameter values choice deployed in the Monte-Carlo simulations.

%\small
\begin{table}[!htbp]
\centering
\caption{System and Channel Parameters considered in the MCS.}
		\label{tab:parameters}
		\vspace{2mm}
		%\footnotesize
		\begin{tabular}{ll}
			\hline
			\bf Parameter & \bf Adopted Value\\
\hline
			Modulation order & $M$ = 2 (BPSK)\\
			SNR & $\gamma \in [0:5:40] $ dB\\
			Number of antennas & $N_t$ = $N_r = N \in [1; 2; 4; 16; 32; 64; 128; 256]$\\
			Adjusting threshold factor & $\rho \in [0.7:0.1:1.3]$ or $\rho \in [0.8:0.05:1.2]$\\
			Steps of Selective-$\rho$ SLAS & $n_F$ = [100; 256; 320]\\
			\hline
						MCS Trials & $\{10^3; \,\, 10^5\}$ realizations\\
						\hline
		\end{tabular}	
\end{table}
%\normalsize

\subsection{BER $\times$ SNR $\times N_t$}

\begin{figure}%[!htbp]
	\centering
	\includegraphics[width=.77515\textwidth]{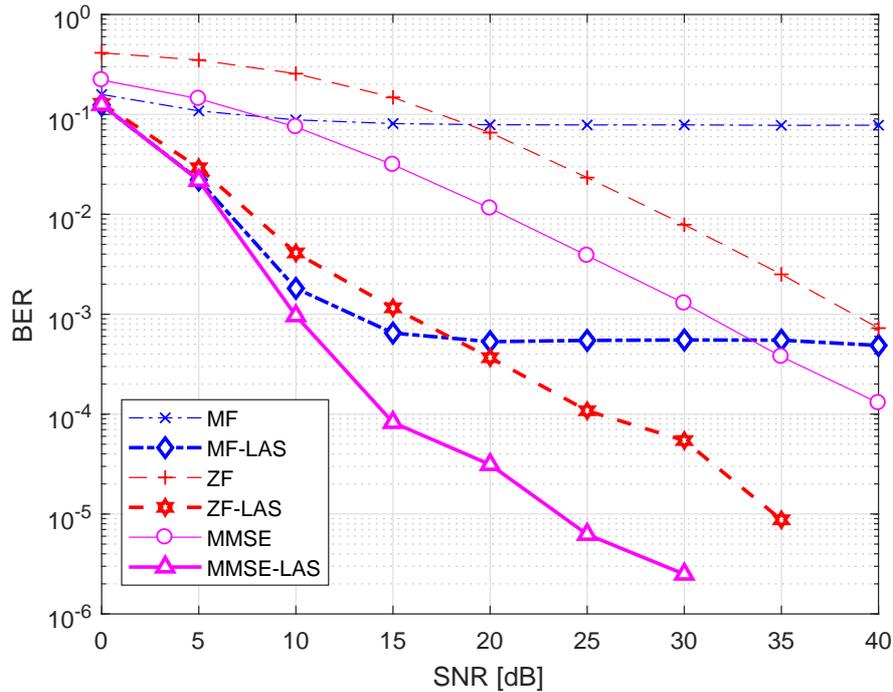}
	\vspace{-5mm}
	\caption{Comparison between linear detectors alone and the optimized version with LAS fixing $N_t= N_r = 32$ antennas}% and changing SNR from 0 to 40 dB.}
	\label{fig:snr}
\end{figure}

Fig. \ref{fig:snr} depicts BER performance of a direct output of the linear detectors (MF, ZF and MMSE) against the optimized output detection with the LAS algorithm for each detector. The SNR was changed from 0 to 40 dB in steps of 5 dB and it was considered 100 steps of LAS and $N_t = N_r = 32$ antennas.

\subsubsection{BER $\times$ SNR Analysis} 
The LAS algorithm never provides a worse solution than the linear detectors alone. For low SNR, the result between the MF and the MF-LAS is very close, however, with the increase of the SNR, it is possible to obtain a BER close to 10$^{-3}$ for high SNR. However, the LAS algorithm can not remove the BER floor due to multiantenna interference in the MF detector. This BER floor can only be mitigated deploying  ZF and MMSE detectors as initial guess for the LAS, but these detectors have a worse performance in low SNR because of noise amplification due to the inverse matrix calculation in such detection process. Hence, combining MMSE or ZF as initial guess with LAS, a BER close to 10$^{-6}$ or 10$^{-5}$ in high SNR can be obtained.

\subsubsection{BER $\times N_t$ Analysis}
\begin{figure}%[!htbp]
	\centering
	\includegraphics[width=.7752\textwidth]{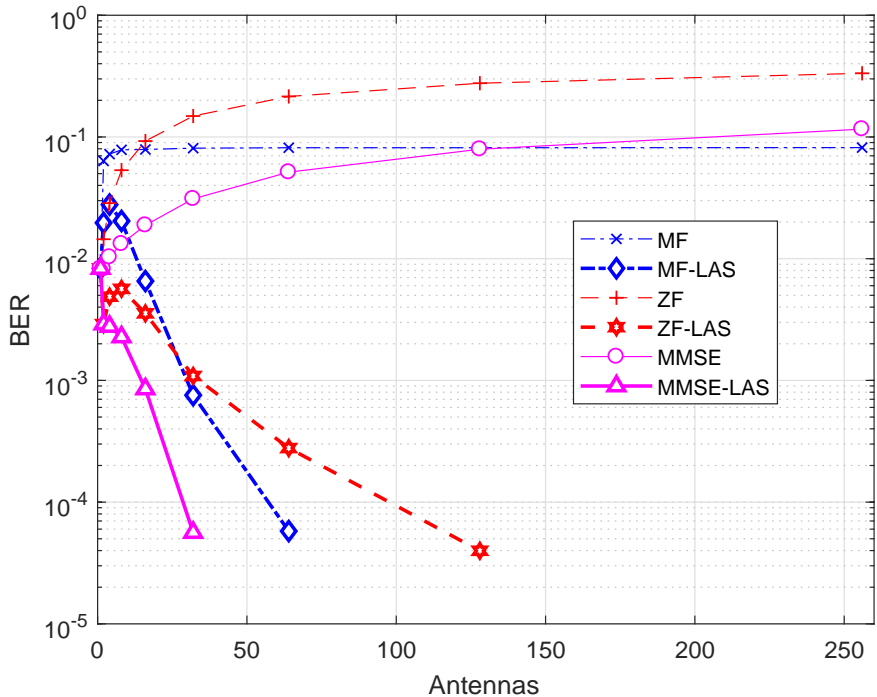}
	\vspace{-6mm}
	\caption{Comparison between linear detectors alone and the optimized version with LAS fixing SNR in 15 dB and changing $N_t$ = $N_r$ from 1 to 256 antennas.}
	\label{fig:antennas}
\end{figure}

In Fig. \ref{fig:antennas}, we simulate a scenario fixing SNR in 15 dB and changing the number of antennas $N_t = N_r$ from 1 to 256 antennas. It was considered 256 steps of LAS algorithm. It is noticed that the performance of linear detectors worsens with the increase of number of antennas. Since the simulated scenario had 15 dB of SNR, there is considerable noise that is amplified in the ZF detector, which makes its performance the worst among the three linear detectors. When applying the LAS algorithm in 256 steps, an improvement in BER performance is observed regarding the increasing number of antennas. As the initial ZF detection was impaired by the low SNR scenario, then this performance propagates to the ZF-LAS as well, showing the worst performance among the three LAS-based MIMO detectors. The MMSE, which results more complexity than the MF, presented better performance in both versions, without and with LAS-aided optimized MIMO detectors. Indeed, under the scenario of Fig. \ref{fig:antennas}, and with $N_t=32$ antennas, the MMSE-LAS detector was able to achieve a BER $\approx 5 \times 10^{-5}$, while the MF-LAS for the same scenario, presented a degradation in performance of approximately one decade, {\it i.e.} BER $\approx 6\times 10^{-4}$.

As can be seen, the linear detectors alone improve the BER due to the interference increase with the number of antennas (since $N_t=N_r$),  until a BER floor be achieved. %with the increase of the number of antennas. 
The ZF detector has a worse performance because with 15 dB of SNR, the noise has a considerable contribution and it is amplified, as can be confirmed by Fig \ref{fig:snr}. Applying LAS, the BER decrease with the increase of number of antennas, showing an inverse performance comparing with the linear detector alone.

\subsection{LAS Convergence}

To verify the LAS convergence to the likelihood performance, we change SNR in low, medium and high levels and observe the behavior of the likelihood function, as well as the BER at each step of the LAS algorithm. For likelihood function analysis, we  have considered MF-LAS, $N_t=128$ antennas, 128 steps of SLAS detector and SNR $\in \{5; \, 10;\, 20\}$ dB.

\begin{figure}%[!htbp]
	\centering
	\includegraphics[width=.7752\textwidth]{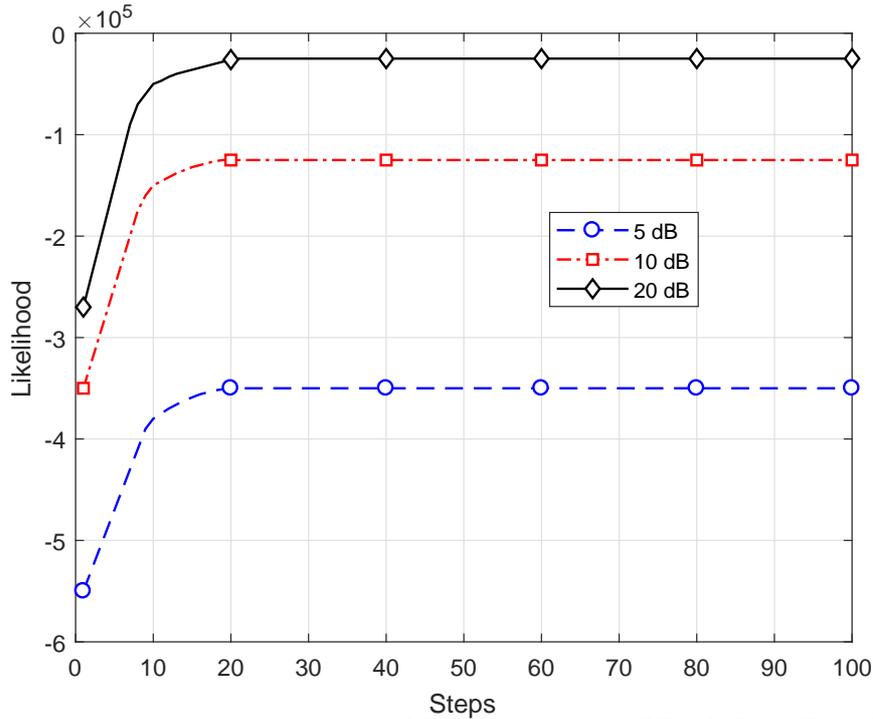}
	\vspace{-7mm}
	\caption{Growth of the likelihood function at each step of the MF-LAS algorithm for SNR = 5 dB, 10 dB and 20 dB.  $N_t=N_r=128$} 
	\label{fig:like}
\end{figure}

Fig. \ref{fig:like} {corroborates} the fact that the LAS achieve performance convergence after $\approx 20$ steps. Indeed, the likelihood function always grows at each step, which is expected from the algorithm, but after $\approx 20$ steps there is no more improvement. It is noted that for $N_t$ = $N_r$ = 128 the algorithm converges to its best solution around 20 steps, indicating the low complexity of the scheme. 

\begin{figure}%[!htbp]
	\centering
	\includegraphics[width=.748\textwidth]{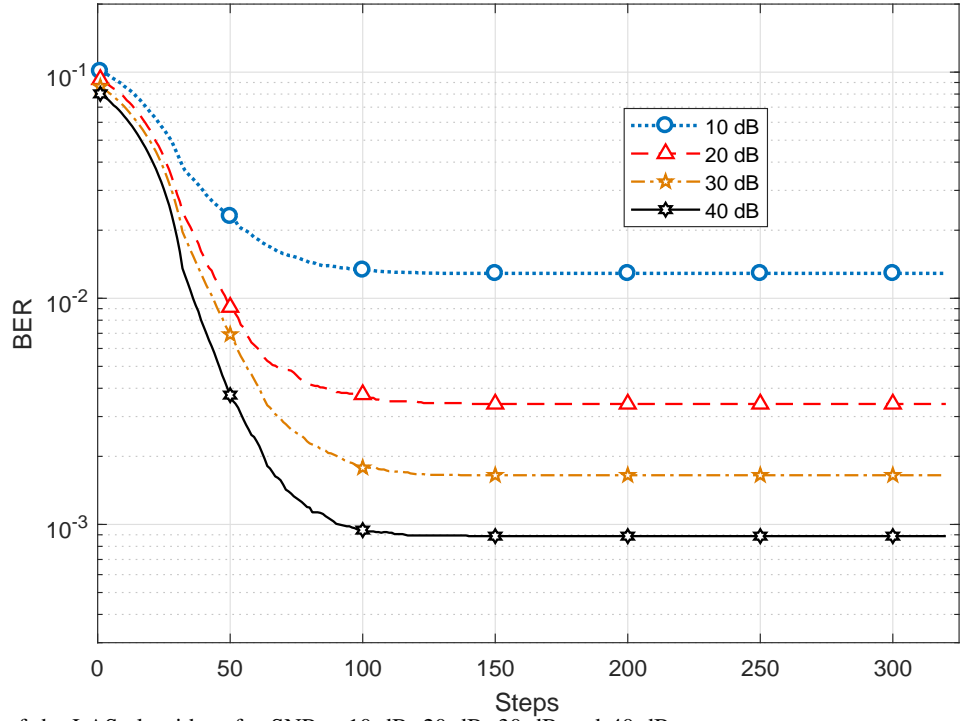}
	\vspace{-7mm}
	\caption{BER of the LAS algorithm, for SNR = 10 dB, 20 dB, 30 dB and 40 dB.}
	\label{fig:ber}
\end{figure}

In terms of BER analysis, Fig. \ref{fig:ber} was obtained with the MF-LAS detector operating under $N_t=64$ antennas, 5 iterations of sequential LAS per antenna (320 steps) and changed the SNR. 
The higher the SNR, the better the MF output, and the BER decreases more abruptly. It can be seen that, about 100 steps, the BER for 40 dB is 5 times smaller then the BER for 10 dB. However, the number of steps for convergence is always approximately 100 steps, not depending on the SNR. However, the BER floor can not be eliminated with LAS algorithm due to the limited strategies available for scape from local optima. On the other hand, the complexity introduced by the sequential LAS optimization step is just marginal.

\subsection{Threshold Analysis}

Changing the threshold $\zeta_j$ from eq. \eqref{eq:threshold} may improve or deteriore the performance of bit flip decisions,  eq. \eqref{eq:rule}. To evaluate possible influence on the BER performance, we have simulated a scenario where the threshold of LAS algorithm is variable, but parameterized in the range $\rho\in [0.7; \;\; 1.3]$.

\begin{figure}%[!htbp]
	\centering
	\includegraphics[width=.752\textwidth]{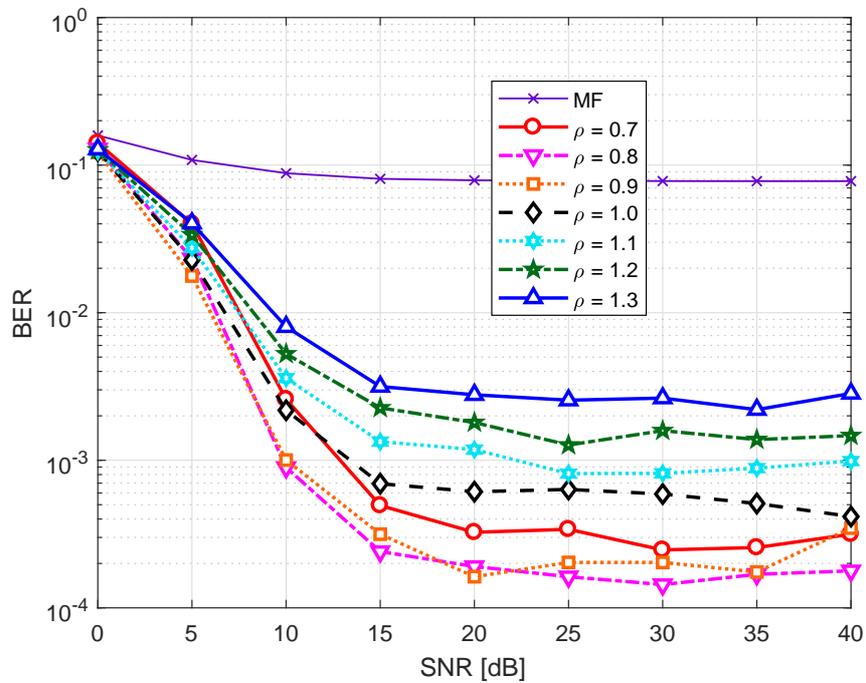}
	\vspace{-6mm}
	\caption{BER analysis of MF-LAS changing the threshold from 70\% to 130\% in steps of 10\%;  $N_t=N_r=32$ antennas.}% 
	\label{fig:threshold}
\end{figure}

Fig. \ref{fig:threshold} shows the BER comparison between the MF detector and the threshold analysis of MF-LAS, changing the $\rho$ from $0.7$ to $1.3$ in steps of $0.1$. It was considered the MF detector for the first decision, $N_t=N_r=32$ antennas and three iterations of sequential LAS per antenna (96 steps).
For a higher threshold, the performance is degraded because of the higher difficulty of bit flipping. For a lower threshold, more bits are flipped and a better solution can be found in local search. Then, the BER is decreased with the decrease of the threshold, but showing an optimal point around $\rho \approx 0.85$ for the considered system configuration scenario.

\begin{figure}%[!htbp]
	\centering
	\includegraphics[width=.752\textwidth]{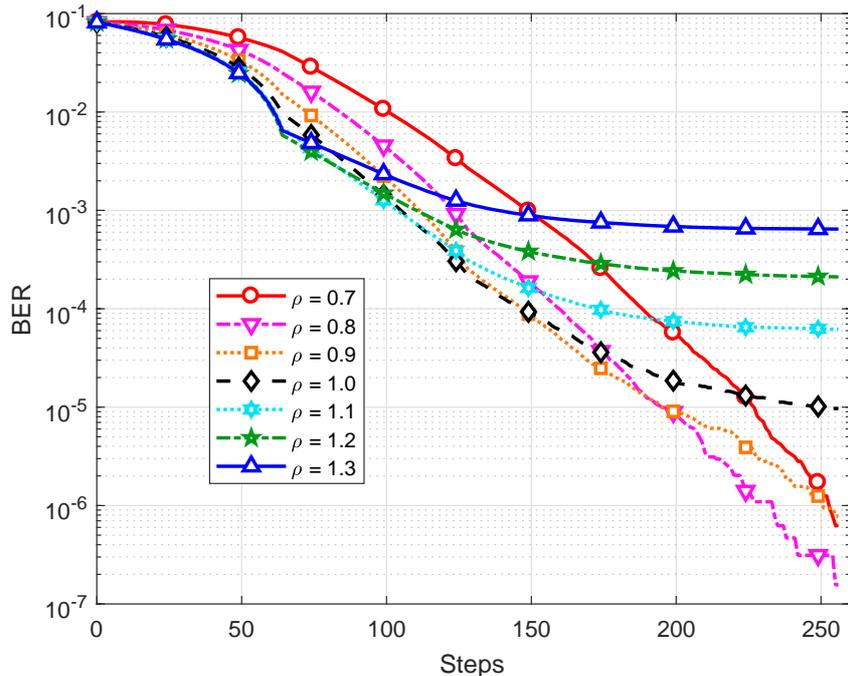}
	\vspace{-7mm}
	\caption{BER per step of MF-LAS algorithm considering threshold analysis. $N_t=N_r=64$ antennas, 4-iter. per antenna (256 steps), and SNR $=15$ dB.}
	\label{fig:threshold_step}
\end{figure}

To verify the change of BER at each step of MF-LAS while {corroborating} our finding regarding the best threshold choice, we have generated Fig. \ref{fig:threshold_step}. In this case, we have considered 64 antennas, four iterations of LAS per antenna (256 steps) and SNR fixed in 15 dB.
As can be seen, a better solution with lower threshold is achieved and it is more quickly and abruptly than the higher threshold. About 250 steps, the normal solution of LAS (100\%) gives a BER of 10$^{-5}$. However, considering a lower threshold, {\it i.e.,} under $\rho=0.8$ (80\% of normal threshold $\zeta_j$), the BER is almost two decades smaller than the $\rho=1.0$ case. Moreover,  considering a higher threshold $\rho=1.3$, the BER is about of $70$ times higher than the normal case. Again, this can be explained because of the bit flipping in the rule of LAS algorithm. For a lower threshold, there are more bit flipping and a better solution can be achieved. However, reducing substantially the threshold level approximate from the random strategy, which is equivalent to implement the conventional MF detection.

\begin{figure}%[!htbp]
\centering
\includegraphics[width=.752\textwidth]{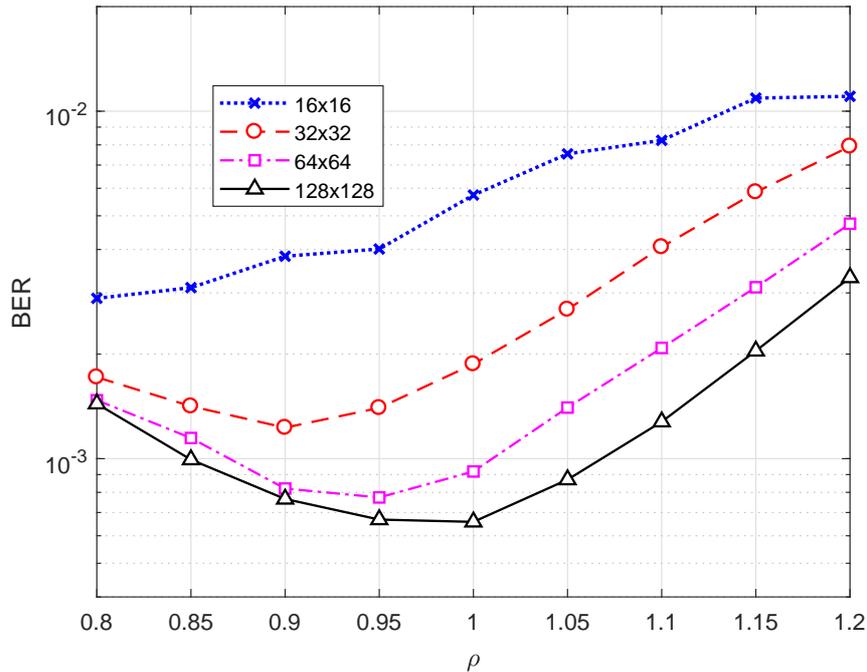}
\vspace{-6mm}
\caption{{BER {\it vs.} $\rho$ for the MF-LAS algorithm considering different number of antennas, 256 steps, and SNR$=10$ dB}.}
\label{fig:berXrho_antenas}
\end{figure}

{For each scenario, there is a $\rho_{\rm opt}$ factor that improves the M-MIMO detector performance. To illustrate this statement, two MCSs were performed plotting the BER as a function of the $\rho$ parameter, which was changed in the range $\rho \in \{0.8:0.05:1.2\}$.} {In the first simulation, the SNR was set at 10 dB and the number of antennas were changed in the range $N \in [16; \;32; \;64; \;128]$. Fig. \ref{fig:berXrho_antenas} shows the results obtained through this MCS. It is observed that the $\rho$ factor that optimizes the BER performance has a tendency to increase with the number of antennas increasing; hence, $\rho_{\rm opt}$ = 0.9 for $32\times32$ and $\rho_{\rm opt}$ = 1.0 for $128\times128$ were found. This means that the optimal bit flipping factor $\rho_{\rm opt}$ has an increasing dependence w.r.t the number of antennas.}
	
\begin{figure}%[!htbp]
\centering
\includegraphics[width=.752\textwidth]{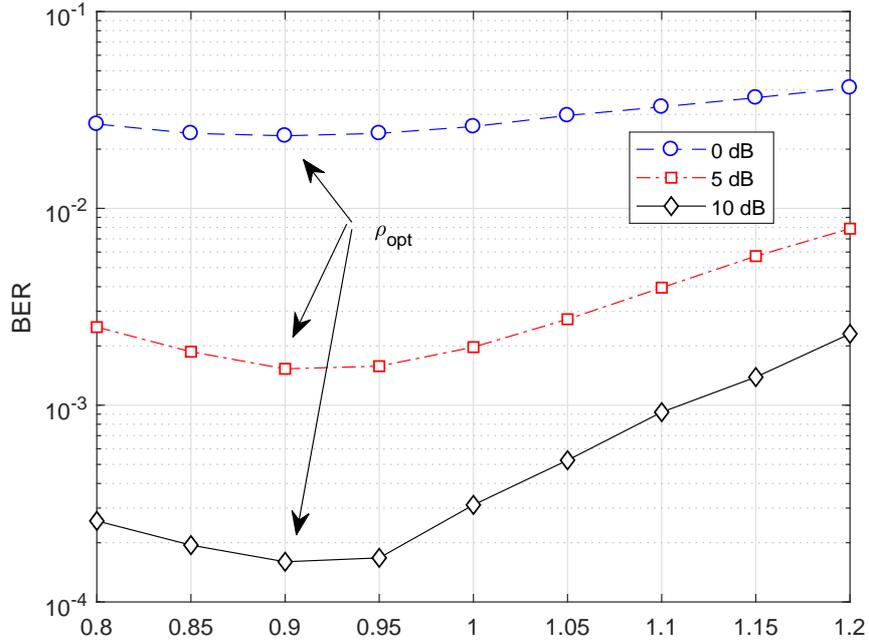}
\vspace{-7mm}
\caption{{BER {\it  vs.} $\rho$ for the MF-LAS algorithm considering 100 steps, and $N$ = 32 antennas.}}
		\label{fig:berXrho_snr2}
	\end{figure}

{In the second scenario, the number of antennas was fixed at $N=32$ and the SNR was adjusted to $\gamma \in [0;\;5;\;10]$ dB values.  Observing  Fig. \ref{fig:berXrho_snr2} one can conclude that for this scenario even with the variation of the SNR, $\rho_{\rm opt}$ tends to 0.9 value. That is, for this scenario, a $\rho_{\rm opt}$ dependence w.r.t SNR has not been observed.}

{For both scenarios, i.e., $N$ or SNR variables there is a $\rho_{\rm opt}$ that optimizes performance. This $\rho_{\rm opt}$ has a strong dependence regarding the number of antennas, while it remains almost unaltered with the SNR variation.}

%------------------------------------------------------------------------------
\section{Complexity Analysis}\label{sec:complexity}
%------------------------------------------------------------------------------

In this section we have carried out a complexity analysis of linear detectors and LAS in the context of massive MIMO detection considering the number of \textit{flops} (floating-point operations) {based on \cite{Matrix}}.

%\subsubsection{Complexity of linear detectors and LAS}
The output of a linear detector such as MF or ZF is used as initial solution and the LAS optimization is applied in a fixed number of steps. As discussed in \cite{Vardhan_LAS08}, the complexity of the ZF-LAS is $ \mathcal{O} (N_tN_r) $, because the complexity of the ZF detector is only $ \mathcal{O} (N_tN_r) $ and the complexity of LAS algorithm is $ \mathcal{O} (N_t) $, so the complexity of the first stage ZF detection is dominant.

To corroborate these complexity reviews, a calculation of the number of flops is performed for each detection process. Considering vectors $\mathbf{x}$ and $\mathbf{y}$ of length $p$. A vector-vector multiplication of type $\mathbf{x}^T\mathbf{y}$ involves $p$ multiplications and $p-1$ additions, totalizing $2p - 1$ flops. In the same way, considering a $m \times p$ matrix $\mathbf{A}$, the matrix-vector multiplication $\mathbf{A}\mathbf{x}$ spends $mp$ multiplications and $m(p-1)$ additions, totalizing $2mp - m$ flops. Also, considering a $p \times n$ matrix $\mathbf{B}$, the matrix-matrix multiplication $\mathbf{A}\mathbf{B}$ involves $mnp$ multiplications and $mn(p-1)$ additions, totalizing $2mnp - mn$ flops. If the variables are complex, then a complex addition spends 2 flops and a complex multiplication spends 6 flops.

We assume that a matrix inversion is made by Gauss elimination, which spends $\frac{2}{3}n^3$ flops of a $n \times n$ matrix \cite{Matrix}, and disregard some computational operations such as allocation, memory access and permutation. Using Eq. \eqref{eq:MF}, \eqref{eq:ZF} and \eqref{eq:MMSE} for the linear detectors and Algorithm \ref{alg:LAS} for the LAS detector, the complexity of each M-MIMO detection technique according to the number of flops to perform data detection is described in Table \ref{tab:complexity}.

\begin{table}[!htbp]
	\centering
	\caption{Complexity of each linear detector and LAS algorithm.}
	\label{tab:complexity}
	\vspace{2mm}
	\footnotesize
	\begin{tabular}{ll}
		\hline
		\bf Detector & \bf Number of Flops\\
		\hline
		MF & $8 N_tN_r - 2N_t$\\
		{ZF} & $\frac{2}{3}N_t^3 + 16N_t^2N_r - 4N_t^2  + 8N_tN_r - 2N_t$\\
		{MMSE} & $\frac{2}{3}N_t^3 + 16N_t^2N_r - 4N_t^2 + 8N_tN_r + 2N_t$\\
		LAS & $8N_t^2n_F$\\
		\hline
	\end{tabular}	
\end{table}

As can be observed, the complexities of ZF and MMSE are very close when the number of antennas tends to be large. Also, the complexity of the LAS algorithm is slightly larger than MF, provided that $N_t \approx N_r$ and that the number of iterations $n_F$ of the algorithm is low. If the algorithm costs many iterations, the complexity of LAS algorithm tends to be larger than the MF, but it is much less complexity cost  regarding the high complexities of the ZF and MMSE detectors.

\begin{figure}%[!h]
\centering
\hspace{-8mm}\includegraphics[width=.752\textwidth]{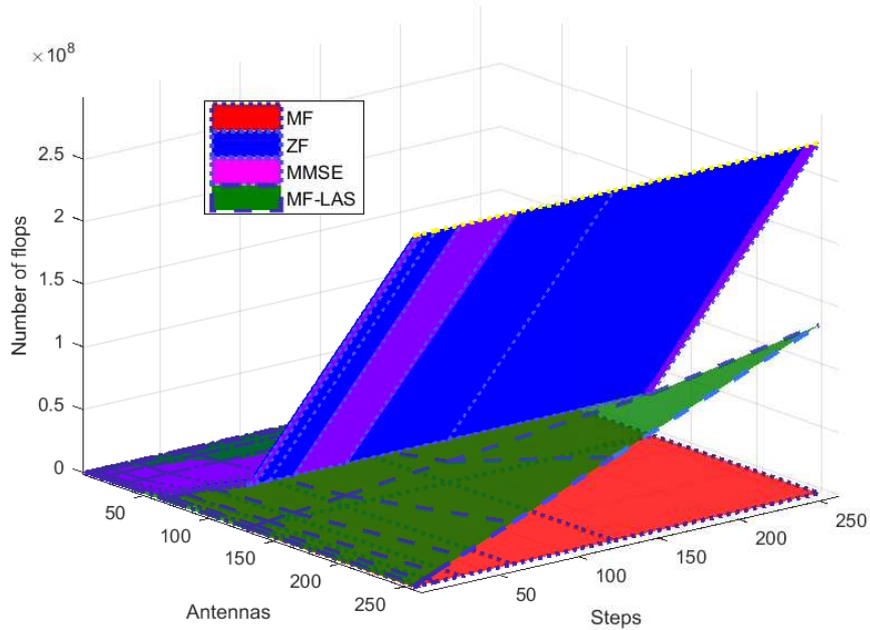}
\vspace{-8mm}
\caption{Number of flops of the detectors considering $N_t = N_r$ and SNR = 0 dB.}
	\label{fig:flops}
\end{figure}

\begin{figure}%[!h]
\centering
\vspace{-10mm}\includegraphics[width=.752\textwidth]{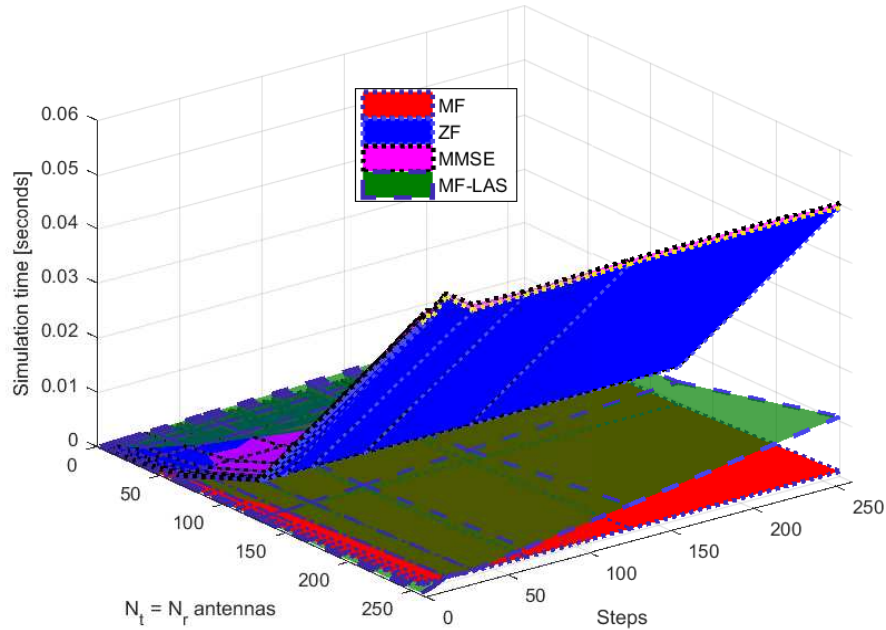}
\vspace{-8mm}
\caption{Simulation time of detectors considering $N_t = N_r$ and SNR = 0 dB.}
	\label{fig:time}
\end{figure}

To demonstrate such complexity orders were have numerically evaluated two  complexity indexes: {\bf a}) number of flops (Fig. \ref{fig:flops}), and {\bf b})  simulation time (Fig. \ref{fig:time}).

Considering the number of flops, the interest was to analyze the behavior of flops according to the variation of the number of antennas and the number of steps of the algorithm LAS. In Fig. \ref{fig:flops}, where $N_t = N_r$ and the SNR was set to 0 dB, which does not greatly change the complexity of the algorithm and the detectors in terms of the number of flops. The number of antennas and the number of steps {were} changed from 1 to 256 in powers of 2, exploring the large scale MIMO scenarios. The ZF and MMSE detectors have very similar complexity, since the surfaces are superimposed. The MF detector presented lower complexity, which confirms what is shown in Table \ref{tab:complexity}. The only curve that changes substantially with the number of steps is the LAS, which presents an inclination in this axis, since the MMSE, ZF and MF detectors do not go include any iterative loop. As the number of steps increases, the increasing in the number of flops of the LAS grows abruptly, reaching approximately half the complexity of the ZF and the MMSE for 256 steps.

In order to complete the analysis, the simulation time is provided in Fig. \ref{fig:time} for the same scenario emulated for the number of flops. It is observed that complexity curves of Fig. \ref{fig:time} closely follows the curves in Fig. \ref{fig:flops}, which indicates that both flop count and time spent are consistent and faithful indexes for the complexity. The small differences can be explained by the simplification of the flop model, which does not take into account the time spent for other computational operations such as memory access, permutation and allocation. Again, the LAS presented a higher simulation time than the MF with the increasing in the number of antennas. The ZF and MMSE detectors, which involve matrix inversion in their processes, presented complexity much higher than the LAS. 

It can be stated that the LAS algorithm does not present high complexity, since its complexity is between the MF that is a low complexity, low performance detector, and the ZF that presents superior performance than MF but is still a linear detector. The process of detecting the initial information has greater weight in complexity determination regarding the LAS steps.

%------------------------------------------------------------------------------
\section{Conclusions}\label{sec:concl}
%------------------------------------------------------------------------------
LAS algorithm for Massive MIMO systems, based on linear MF, ZF and MMSE detectors as the initial guess have been extensively analyzed. The BER performance comparison for different M-MIMO system scenarios was carried out, and we have found that using an adjustable (lower than that found in the literature) threshold factor for the bit flipping rule of LAS, a better solution can be achieved in terms of BER. {It also indicates that an optimum threshold can be found in function of the number of antennas and SNR.}

Our analysis revealed  that the threshold optimization of the LAS algorithm increases substantially the performance when SNR, the number of transmitting and receiving antennas are increased. The MF-LAS do not cancel the multiantenna interference of the MF detector. As a result, the BER of MF-LAS is higher than ZF-LAS and MMSE-LAS with the increase of SNR. However, the ZF and MMSE detector {need} to compute a matrix inversion, increasing complexity.

The complexity of LAS detector algorithm for M-MIMO is higher than the MF, which depends on the number of steps in the algorithm. However, the LAS {has} resulted in lower complexity than the linear detectors ZF and MMSE, which indicates that the greater weight of the detection is in obtaining the initial solution, since the steps of the algorithm present less complexity than the linear detectors ZF and MMSE, those who are known to have {high complexity in M-MIMO systems due to the matrix inversion}.

\section*{Acknowledgement}
This work was supported in part by the National Council for Scientific and Technological Development (CNPq) of Brazil under Grants 304066/2015-0, and in part by CAPES - Coordenacao de Aperfeicoamento de Pessoal de N­ivel Superior, Brazil (scholarship),  and by the Londrina State University - Parana State Government (UEL).

\end{document}